\documentclass[prd,aps,psfig,floats]{revtex4}

\usepackage{graphicx}
\usepackage{dcolumn}
\usepackage{amsmath}

\begin{document}

\title{On Perturbations in Warm Inflation}

\author{H. P. De Oliveira}
 \email{henrique@fnal.gov}
\affiliation{{Nasa/Fermilab Astrophysics Center \\ 
Fermi National Accelerator Laboratory, Batavia, Illinois, 60510-500.}\\ 
And \\
{\it Universidade Do Estado Do Rio De Janeiro }\\ 
{\it Instituto De F\'{\i}sica - Departamento De F\'{\i}sica Te\'orica,}\\ 
{\it Cep 20550-013 Rio De Janeiro, RJ, Brazil}}

\author{S. E. Jor\'as}
\email{joras@het.brown.edu}
\affiliation{Department Of Physics, Brown University, Providence, RI 02912, USA}

\date{\today}

\begin{abstract} 

Warm inflation is an interesting possibility of describing the early universe,
whose basic feature is the absence, at least in principle, of a preheating or
reheating phase. Here we analyze the dynamics of warm inflation generalizing
the usual slow-roll parameters that are useful for characterizing the
inflationary phase. We study the evolution of entropy and adiabatic
perturbations, where the main result is that for a very small amount of
dissipation the entropy perturbations can be neglected and the purely
adiabatic perturbations will be responsible for the primordial spectrum of
inhomogeneities. Taking into account the COBE-DMR data of the  cosmic 
microwave
background anisotropy as well as the fact that the
interval of inflation for which the scales of astrophysical interest cross
outside the Hubble radius is about 50 e-folds before the
end of inflation, we could estimate the magnitude of the dissipation term. It
was also possible to show that at the end of inflation the universe is hot
enough to provide a smooth transition to the radiation era.

\end{abstract}

\pacs{PACS number:xxx}

\maketitle

\newpage

\section{Introduction}

An inflationary phase gets rid of the major problems in cosmology, 
namely flatness, the horizon problem, homogeneity and numerical density of
monopoles\cite{kolbtur}. As strange as it may sound at first, it is actually
very easy to have an exponential behavior of the scale factor. The main problem
is how to attach the observed universe to the end of the inflationary epoch. 

There are three possible solutions to this problem: reheating\cite{kolbtur}, 
preheating\cite{preheat}, and warm inflation\cite{berefang,olivrud,gleiser,
bellini} (WI, from now on). In this work we focus on the latter, according 
to which the radiation is produced throughout inflation such that its energy
 density  is kept nearly constant in this phase. This is accomplished by 
 introducing adissipation term, $\Gamma$, in the equation of motion for the 
 inflaton, granting a continuous energy transfer from its decay. As a 
 consequence, no reheating or preheating is necessary since we  expect that 
 enough radiation is produced to provide a smooth transition to the radiation 
 era at the end of inflation. 
	
On the other hand, being a two-field model, we are obliged to take into account
entropy (isocurvature) perturbations\cite{entropy,bardeen} if the problem
of structure formation 
is considered. As is well known, the very existence of the former may provide
non trivial evolution of the curvature perturbation on comoving hypersurfaces,
$\mathcal{R}$, since this quantity  is no longer frozen for large
scale modes as in single field inflationary models. The usual procedure for
determining the amplitude of the observed perturbations takes the fact that
$\dot\mathcal{R}=0$ for granted as soon as a given mode crosses outside the
Hubble radius during inflation. The dissipation term is of central importance
for the entropy perturbations, since in its absence the radiation  is
decoupled from the scalar field and therefore redshifted away together with its
fluctuations during inflation. Then the overall scenario is reduced to a
single field, where it is known that the entropy perturbations are neglected
for long wavelength modes. In the presence of dissipation,
radiation is rather produced than washed away and the  effective fluid is a 
mixture
of mutually interacting radiation and scalar field. The
variation of its own effective equation of state is responsible for the entropy 
perturbations.

The main goal of this paper is to discuss the problem of perturbations in WI;
the role of the dissipation term in the evolution of the entropy perturbation is
exhibited after the development of a consistent gauge-invariant
procedure. 

The paper is organized as follows. In Section II the basic equaitons of WI are
introduced together with a generalization of the slow-roll parameters that
allows a better charaterization of the slow-roll phase and its end. The mixture
of radiation and scalar field can be described as an effective fluid with a
well defined equation of state. In Section III a consistent gauge-invariant
treatment is developed in which the connection between entropy perturbations
and dissipation is established. The expressions for scalar adiabatic 
perturbations
and its spectral index in WI are shown in Section IV. Section V is devoted to
presenting two models in WI, both in the weak dissipation regime. In this
case the entropy perturbations can be neglected, and an estimate of the
magnitude of the dissipation term was obtained, together with other relevant
parameters of the model, after taking into account the COBE data. Although the
smallness of the dissipation, the temperature at the end of inflation was
calculated and shown to be reasonable for a smooth transition to the radiation
era. Finally, in Section VI we conclude and trace out some perspectives of the
present work.

\section{Dynamical aspects of warm inflation and the generalized  slow-roll
parameters}

Berera and Fang\cite{berefang} considered flat FRW models endowed with 
interacting radiation and scalar field whose dynamics is governed by the 
Friedmann equation

\begin{equation}
H^2=\frac{8\pi}{3m^2_{pl}} 
\left( \rho_r +\frac{1}{2}{\dot\phi}^2+V(\phi)\right)\;,
\label{eq1}
\end{equation}
by the equation of motion of an homogeneous scalar field $\phi(t)$ in the 
effective potential $V(\phi)$
\begin{equation}
\ddot \phi +(3H+\Gamma)\dot \phi + V^\prime[\phi(t)]=0\;,
\label{eq2}
\end{equation}
and by the $1^{st}$ Law of Thermodynamics
\begin{equation}
\dot \rho_r + 4 H \rho_r = \Gamma {\dot\phi}^2\;.
\label{eq3}
\end{equation}
In these equations $H=\frac{\dot{a}}{a}$ is the Hubble parameter, $m_{pl}$ is 
the Planck mass, dot and prime denote derivative with respect to the 
cosmological time $t$ and the
scalar field, respectively; $\rho_r$ is the energy density of the radiation
field. The friction term $\Gamma$ is responsible for the decay of
the scalar field into radiation. 

Warm inflation is characterized by the accelerated growth of the scale factor
driven by the potential term $V(\phi)$ that dominates over other energy terms in
Eq. (\ref{eq1}). During this phase: (i) the Friedmann equation becomes
\begin{equation}
H^2 \simeq \frac{8\pi}{3m^2_{pl}} V(\phi)\; ;
\label{eq4} 
\end{equation} 
(ii) the $\ddot{\phi}$ term is neglected in Eq. (\ref{eq2}), yielding 
\begin{equation}
\dot{\phi} \simeq -\frac{V^{\prime}(\phi)}{3 H + \Gamma}\; ;
\label{eq5}
\end{equation}
(iii) and finally, since the scalar field steadily decays into radiation, it is 
assumed that $\dot{\rho_r} \approx 0$, or equivalently,
\begin{equation}
\rho_r \simeq \frac{\Gamma \dot{\phi}^2}{4 H}
\label{eq6}
\end{equation}

The inflationary regime is more precisely characterized by the introduction of
the slow roll parameters. For the present scenario the generalization of the
usual slow-roll parameters is given by

\begin{equation}
\epsilon_{wi} = \frac{3 \dot{\phi}^2 + 4 \rho_r}{2 (V(\phi)+\frac{1}{2} 
\dot{\phi}^2 + \rho_r)} = \frac{m^2_{pl}}{4 \pi (1+\alpha)}
\left(\frac{H^{\prime}(\phi)}{H(\phi)}\right)^2 - \frac{1}{3 (1+\alpha)}
\frac{H^{\prime} \rho_r^{\prime}}{H^3}
\label{eq7}
\end{equation}
and
\begin{equation}
\eta_{wi} = -\frac{\ddot{\phi}}{H \dot{\phi}} =  
\frac{m^2_{pl}}{4 \pi (1+\alpha)} \frac{H^{\prime\prime}(\phi)}{H(\phi)} -
\frac{\alpha^{\prime} H}{(1+\alpha) H^{\prime}} \epsilon_{wi} - 
\frac{1}{3 (1+\alpha) H}\,\left(\frac{\rho_r^{\prime}}{H}\right)^{\prime}, 
\label{eq8}
\end{equation}

\noindent where we have introduced $\alpha \equiv \frac{\Gamma}{3 H}$ that can
be either greater or less than the unity depending on the regime of strong or
weak dissipation. Note that the above expressions are exact. The slow-roll phase 
is characterized by $\epsilon_{wi} \ll 1$, as a consequence of $V(\phi) \gg 1/2
\dot{\phi}^2+\rho_r$, which is necessary to validate Eq. (\ref{eq4}), and by 
$\eta_{wi} \ll 1$ for the condition of neglecting the $\ddot{\phi}$ term in
Eq. (\ref{eq2}). Besides these conditions we must require $\rho_r^{\prime}
\simeq 0$ that stands for the quasi-stable production of radiation. During the
slow-roll, these parameters can be expressed in terms  of the usual slow-roll
parameters $\epsilon$ and $\eta$ of the super-cooled inflation as 
\begin{equation}
\epsilon_{wi} \simeq \frac{\epsilon}{1+\alpha}
\label{eq9}
\end{equation}
and
\begin{equation}
\eta_{wi} \simeq \frac{\eta}{1+\alpha} + \frac{\alpha^{\prime} H}{(1+\alpha)^2
H^{\prime}} \epsilon,
\label{eq10}
\end{equation}

\noindent 
where $\epsilon \equiv \frac{m^2_{pl}}{4 \pi}
\left(\frac{H^{\prime}(\phi)}{H(\phi)}\right)^2$ and 
$\eta \equiv  \frac{m^2_{pl}}{4 \pi}\left(\frac{H^{\prime\prime}(\phi)}
{H(\phi)}\right)$ as usually defined\cite{parinflation}.

The parameters $\epsilon_{wi}$ and $\eta_{wi}$ are natural and direct
generalizations of the parameters $\epsilon$ and $\eta$ of the super-cooled
inflation. In the absence of dissipation and radiation, the
parameters defined in warm inflation reduce to the usual ones. Besides, the 
parameter $\epsilon_{wi}$ can also be understood as a direct
measure of the effective equation of state relating the total pressure
$p=\frac{1}{2} \dot{\phi}^2 - V(\phi) + p_r$ and the
total energy $\rho = \frac{1}{2} \dot{\phi}^2 + V(\phi) + \rho_r$ through
$p=-\left(1-\frac{2}{3} \epsilon_{wi}\right) \rho$. The definition of 
inflation as the
period of accelerated expansion of the universe implies that
$\frac{\ddot{a}}{a}>0$, so that from the exact expression

\begin{equation}
\frac{\ddot{a}}{a} = (1-\epsilon_{wi}) H^2
\label{eq11}
\end{equation} 

\noindent the  inflationary epoch can be formally characterized by
$\epsilon_{wi}<1$. The end of inflation occurs for $\epsilon_{wi}=1$, implying
that, in this stage $\frac{1}{2} \dot{\phi}^2+\rho_r$ is comparable
to the potential $V(\phi)$, or equivalently
$\left(\frac{1}{2} \dot{\phi}^2+\rho_r\right)_{end}= V(\phi_{e})$ (see
Eq. (\ref{eq7})), where $\phi_{e}$ is the scalar field at the end of
inflation. The condition $\dot{\rho}_r \simeq 0$ is valid in most of inflation,
but it is violated at the beginning and at the end of inflation, as well.

The crucial aspect of warm inflation is that the temperature of radiation,
$T_r$, must exceed the Hawking temperature, $H$, or 

\begin{equation}
T_r \geq H,
\label{eq12}
\end{equation}

\noindent
during the inflationary expansion; the moment when $T_r=H$ is
considered the beginning of warm inflation. This relation assures that thermal
fluctuations of the scalar field dominate over the quantum fluctuations
producing a distinct spectrum of primordial fluctuations from that of
super-cooled inflation\cite{berefang,leefang1}.

Finally, we derive a very useful expression for the number of e-folds $N$
associated to the scalar field. We set $a=a_{e} e^{N(\phi)}$, where
$a_{e}=a(\phi_{e})$, $\phi_{e}$ are the values of the scale factor and
the scalar field at the end of inflation, $N$ is the number of e-folds before
the end of inflation. It can be easily found that, in the slow-roll 
approximation,
 
\begin{equation}
N  = \pm \frac{4 \pi}{m^2_{pl}} \int^{\phi_{e}}_{\phi_{i}}{(1+\alpha)
\frac{H(\phi)}{H^{\prime}(\phi)} d \phi} 
\label{eq13}
\end{equation}

\noindent 
where the upper (lower) sign is used when $\dot{\phi}>0(<0)$, so that $N$ is 
always a positive quantity.
Since we know that warm
inflation is characterized by the dominance of the thermal component of
radiation, it begins when $T_r=H$ at $\phi=\phi_*$, say. Then,
denoting by $N_{wi}$ the number of e-folds during the warm inflation, we have

\begin{equation}
N_{wi}  = \pm \frac{4 \pi}{m^2_{pl}} \int^{\phi_{e}}_{\phi_{*}}{(1+\alpha)
\frac{H(\phi)}{H^{\prime}(\phi)} d \phi}.
\label{eq14}
\end{equation}

\noindent 
In general, $\phi_* \neq \phi_i$ and $\phi_*$ depends on the typical
parameters of the theory.

\section{Perturbations in warm inflation}

We consider the inhomogeneous perturbations of the FRW background described by
the metric in the longitudinal gauge\cite{entropy,bardeen} 

\begin{equation}
d s^2 = (1+2 \Phi)\,d t^2 - a^2(t) (1-2 \Psi) \delta_{i j}\,d x^i\,d x^j,
\label{eq15}
\end{equation}

\noindent where $a(t)$ is the scale factor, $\Phi=\Phi(t,\bf{x})$ and
$\Psi=\Psi(t,\bf{x})$ are the metric perturbations. The  background matter
content is constituted by radiation and scalar field interacting through the
friction term $\Gamma$ as shown in the last Section. The spatial dependence of all
perturbed quantities are of the form of plane waves $e^{i\bf{k.x}}$, $k$
being the wave number, so that the perturbed field equations regarding now only
their temporal parts (we omit the subscript $k$) are

\begin{eqnarray}
-3 H (H \Phi + \dot{\Phi}) - \frac{k^2}{a^2} \Phi = \frac{4 \pi}{m_{pl}^2}
\delta \rho \nonumber \\   
\dot{\Phi} + H \Phi = \frac{4 \pi}{m_{pl}^2} \left(-\frac{4}{3 k} \rho_r a v
 + \dot{\phi}_0 \delta \phi \right) \label{eq16} \\
 \ddot{\Phi} + 4 H \dot{\Phi} + (2 \dot{H} + 3 H^2) \Phi = \frac{4 \pi}{m_{pl}^2}
\delta p.  
\nonumber
\end{eqnarray}

\noindent In the above equations dot means derivative with respect to $t$,
$\rho_r$ and $\phi_0$ are background energy 
density of radiation and the scalar field, respectively; $\delta \rho = \delta
\rho_r + \dot{\phi_0}\,(\delta \phi \dot{)} - \dot{\phi_0}^2\,\Phi +
V^{\prime}\,\delta \phi$ and $\delta p = \frac{1}{3} \delta \rho_r +
\dot{\phi_0}\,(\delta \phi \dot{)} - 
\dot{\phi_0}^2\,\Phi - V^{\prime}\,\delta \phi$ are the perturbations of the
total energy density and pressure, respectively; $v$ originates from the
decomposition of the velocity field as $\delta U_i=-\frac{i a
k_i}{k}\,v\,e^{i \bf{k.x}}$ (see Bardeen\cite{bardeen}). Also, due the fact
that the perturbation of the total energy-momentum tensor does not give rise to
anisotropic stress ($\delta T^i_j \propto \delta^i_j$), $\Psi=\Phi$. For the
sake of completeness we present in the appendix the evolution equations for
$\delta \phi$, $\delta \rho_r$ and $v$.

We assume that $\delta p$ and $\delta \rho$ are connected by the relation

\begin{equation}
\delta p = c_s^2\,\delta \rho + \tau\,\delta S
\label{eq17}
\end{equation} 

\noindent where $c_s^2 =
\frac{\dot{p}}{\dot{\rho}}=\frac{\dot{p}_r +
\dot{p}_{\phi}}{\dot{\rho}_r+\dot{\rho}_{\phi}}$ 
is the effective velocity of sound of the fluid constituted by the mixture of
radiation and scalar field. The quantity $\tau\,\delta S$ is the contribution
to the perturbation of the pressure due to the variation of the effective
equation of state that relates $p$ and $\rho$, or the entropy perturbation. 
Introducing Eq. (\ref{eq17}) into the third equation of the system
(\ref{eq16}), and taking into account the first equation, we obtain

\begin{equation}
\ddot{\Phi} + (4+3 c_s^2)\,H\,\dot{\Phi} + [2 \dot{H} + 3 H^2\,(1+c_s^2)]\,\Phi
+ \frac{c_s^2 k^2}{a^2}\,\Phi = \frac{4 \pi}{m^2_{pl}}\,\tau \delta S.
\label{eq18}
\end{equation}

\noindent The rhs of the above equation accounts for the entropy
perturbations. In the absence of such perturbations and for long wavelength
perturbations, the gauge-invariant curvature perturbation on comoving
hypersurfaces\cite{entropy,gordon}

\begin{equation}
\mathcal{R} = \frac{2\,(H \Phi + \dot{\Phi})}{3 H\,(1+w)} + \Phi
\label{eq19}
\end{equation}

\noindent with $w = \frac{p}{\rho} = \frac{p_r+p_{\rho}}{\rho_r+\rho_{\phi}}$,
is conserved. In the absence of the scalar field $c_s^2=w=\frac{1}{3}$, $\tau
\delta S = 0$ (see also next expression). Writing $
\rho = \rho_r + \frac{1}{2}{\dot{\phi}}^2 + V(\phi_0)$ and $p =
\frac{1}{3}\rho_r + \frac{1}{2}{\dot{\phi}}^2 - V(\phi_0)$ in Eq. (\ref{eq17}),
yields

\begin{equation}
\tau \delta S = (1-c_s^2)\,\delta \rho -\frac{2}{3} \delta \rho_r - 2
V^{\prime}\,\delta \phi 
\end{equation}

\noindent After some manipulation using the field equations (\ref{eq16}), it 
results in 
\begin{equation}
\tau \delta S = -\frac{m_{pl}^2}{4 \pi}\,({H \Phi + \dot{\Phi}})\,A -
\frac{m_{pl}^2 (1-c_s^2)}{4 \pi a^2}\,k^2\,\Phi -
\frac{2}{3}\,\rho_r\,\left(\frac{4 a V^{\prime} v}{k \dot{\phi}_0} + \frac{\delta
\rho_r}{\rho_r}\right) 
\label{eq20}
\end{equation} 

\noindent where $A$ is given by

\begin{equation}
A = \frac{8 \rho_r\,(H+V^{\prime}/\dot{\phi}_0)-2 \Gamma
\dot{\phi}^2_0}{3\,(\dot{\phi}^2_0+4/3 \rho_r)}.
\label{eq21}
\end{equation}

\noindent  Assuming slow-roll conditions (\ref{eq4})-(\ref{eq6}), it can be
shown that $A$ is given by $A \approx -2 \Gamma$, putting in evidence the role
of dissipation in the entropy perturbations.
\noindent  
In the case of $\Gamma = 0$ the radiation field decays
exponentially during inflation implying that it can be neglected, and therefore
the entropy perturbation term is reduced to
\begin{equation}
\tau \delta S=\frac{m_{pl}^2
(1-c_s^2)}{4 \pi}\,\frac{k^2\,\Phi}{a^2}=\frac{m_{pl}^2 V^{\prime}}{6 \pi H
\dot{\phi}_0}\,\frac{k^2\,\Phi}{a^2}\;,
\end{equation}
which agrees with previous results\cite{gordon} obtained for a single 
scalar field, and that vanishes for long wavelength perturbations. In this way,
the dissipation term plays a crucial role in producing entropy perturbations by
preventing the radiation field field to be washed away during
inflation. Therefore, during WI different matter components, scalar field and
radiation, evolve such that eventually nonuniform spatial distributions but
with uniform total energy density are present. This is, roughly speaking, the
origin of the entropy perturbations.

Now, we can rewrite Eq. (\ref{eq19}) a very useful form considering long
wavelength perturbations. Using the comoving curvature perturbation
$\mathcal{R}$, we have

\begin{equation}
\dot{\mathcal{R}} = -\frac{2\,(H \Phi + \dot{\Phi})\,A}{3 H\,(1+w)} - \frac{16
\pi \rho_r}{9 m_{pl}^2 (1+w) H}\left(\frac{4 a V^{\prime} v}{k \dot{\phi}_0} +
\frac{\delta \rho_r}{\rho_r}\right)
\label{eq22}
\end{equation}   
                                          
\noindent which relates the change in the comoving curvature perturbation due
to the source $\tau \delta S$. We can determine the dominant term of the rhs of
the above equation during inflation (see appendix for details), that shows to
be  

\begin{equation}
\dot{\mathcal{R}} \simeq \frac{4 \rho \Gamma}{3\,(\rho+p)}\,\Phi
\label{eq23}
\end{equation}

\noindent As expected from the previous discussion, the entropy perturbation
depends directly on the dissipation term. For very small dissipation, which is
still allowed in warm inflation, the source term in the above equation can be
neglected and the adiabatic perturbations will be mostly responsible for the
primordial spectrum of inhomogeneities. Independently of all approximations
used it is intuitive that the dissipation may play a central role in producing
entropy perturbations, since in the absence of dissipation the radiation field is
washed away during inflation and its effect in the overall scenario can be
neglected. We remark that our result is in agreement with Taylor and
Berera\cite{taylor}, although a gauge-invariant analysis were not carried out,
and opposite to the one obtained by Wolung Lee and Fang\cite{leefang2}.

In the last section we will consider a small dissipation such that
Eq. (\ref{eq23}) can be reduced to $\dot{\mathcal{R}} \simeq 0$, meaning that
the entropy perturbations are no longer important and the primordial spectrum of
perturbations is due only to adiabatic perturbations.

\section{Spectral index for scalar perturbations}

Based on the final discussion of the last section the spectrum of scalar
adiabatic perturbations is given by

\begin{equation}
\delta_{ad} \simeq \frac{\delta \rho}{\rho+p} = \frac{3
H}{\dot{\phi}_0}\,\delta \phi. 
\label{eq24}
\end{equation}   
      
\noindent This expression must be evaluated for $k = a H$, that is when a given
scale $k$ crosses the Hubble radius. The fluctuations of the scalar field are now
supposed to be of thermal origin instead of quantum fluctuations, or
equivalently\cite{berefang,leefang1} 

\begin{equation}
(\delta \phi)^2 \simeq \frac{3 H}{4 \pi}\,T_r
\label{eq25}
\end{equation}

\noindent where $T_r$ is the temperature of the thermal bath. For quantum
fluctuations we know that $(\delta \phi)^2_{quant} \approx \frac{H^2}{4 \pi}$,
so that the dominance of thermal fluctuations is guaranteed for $T_r > H$ which
occurs during the warm inflation. Despite the small dissipation, the spectrum
of perturbations will depend strongly on the dissipation as we can see after
substituting Eq. (\ref{eq25}) into Eq. (\ref{eq24}), and taking into account that

\begin{equation}
T_r = \left(\frac{30}{g \pi^2}\,\rho_r\right)^{1/4} \simeq \left(\frac{15}{2
\pi^2 g}\,\frac{\Gamma \dot{\phi}^2_0}{H}\right)^{1/4}
\label{eq26}
\end{equation}
      
\noindent which holds during the slow-roll, with $g \approx 100$ being the
number of degrees of freedom for the radiation field. Then, we obtain

\begin{equation}
\delta_{ad}^2 \simeq
37.19\,\left[\frac{\alpha^{1/6}\,(1+\alpha)}{H^{\prime}}\right]^{3/2}\, 
\left(\frac{H}{m_{ pl}}\right)^3
\label{eq27}
\end{equation}

It is also useful to describe this spectrum in terms of its spectral
index, $n_s$, defined as 

\begin{equation}
n_s - 1 = \frac{d\,ln\,\delta_{ad}^2}{d\,ln\,k}.
\label{eq28}
\end{equation} 

\noindent Using the usual assumptions (i) $a=e^{-N}\,a_{e}$ (see Section II);
(ii) the fact that $k=a\,H$ for a given scale crossing the Hubble radius, one
can show that  

\begin{equation}
\frac{d\,ln\,k}{d\,\phi_0} = \frac{4
\pi}{m^2_{pl}}\,\frac{H}{H^{\prime}}\,(1+\alpha)\,(\epsilon_{wi}-1). 
\label{eq29}
\end{equation} 

\noindent For $\alpha = 0$, it follows that $\epsilon_{wi} = \epsilon$ and the
expression established in super-cooled inflation is easily
recovered\cite{kolbdodel}. Finally, considering Eqs. (\ref{eq27}), (\ref{eq28})
and (\ref{eq29}) we found the following expression for $n_s$ in terms of the
slow-roll parameters 

\begin{equation}
n_s = 1 - \frac{1}{2 (1 - \epsilon + \alpha)}\,\left(\frac{(11 + 5
\alpha)\,\epsilon}{2\,(1 + \alpha)} - 3 \eta \right) - \frac{m^2_{pl} (1 + 7
\alpha)\,\Gamma^{\prime} H^{\prime}}{48 \pi \alpha\,(1 + \alpha)\,(1 - \epsilon
+ \alpha)\,H^2}
\label{eq30}
\end{equation}

\noindent We remark that the above equation is general in the sense that no
approximation regarding $\alpha$ was used, and there is a contribution of the
variation of the dissipation term that can be relevant. For instance,
let us consider the potential of the type $V(\phi) = \lambda^4 \phi^q$, and a
constant dissipation $\Gamma_0$. It follows that for
$\alpha \ll 1$, $n_s 
\simeq 1 - \frac{1}{2}\,\left(\frac{11}{2} \epsilon - 3 \eta\right)$, and for
any $q$, $n_s$ is always less than one as in super-cooled inflation. On the
other hand, if $\alpha \gg 1$, $n_s \simeq 1 - 
\frac{1}{2\alpha}\,\left(\frac{5}{2} \epsilon - 3 \eta\right)$, where one 
can easily show that only if $q>12$ the index $n_s$ can be greater than one 
giving rise to a blue spectrum. In the case of variable dissipation, we assume 
$\Gamma = \beta_2 \phi^2$ and the corresponding expression 
for the index $n_s$ becomes $n_s \simeq 1 - \frac{1}{2 (1+\alpha-\epsilon)}\,
\left(\frac{(11 q + 4) + (5 q + 28) \alpha}{2 q (1+\alpha)}\,\epsilon - 
3 \eta\right)$. In particular, if $q = 4$ this expression is reduced to $n_s
\simeq 1 - \frac{3}{2 (1+\alpha - \epsilon)}\,(2 \epsilon - \eta)$ which is
always less than 1.

\section{Some worked examples}

In order to estimate consistently the role of dissipation during the warm
inflation, let us consider the potential

\begin{equation}
V(\phi) = \lambda^4\,\phi^4
\label{eq31}
\end{equation}

\noindent where $\lambda$ is an adimensional parameter. According to
ref. \cite{olivrud}, the dissipation term can be written as $\Gamma =
\beta_m\,\phi^m$, $m=0,2$. For the sake of simplicity we set $m=2$ which yields

\begin{equation}
\alpha = \frac{m_{pl} \beta_2}{\sqrt{24 \pi}\,\lambda^2},
\label{eq32}
\end{equation}

\noindent that is a constant, and $[\beta] = m_{pl}^{-1}$. During inflation
Eq. (4) holds, and together with Eq. (5) the density of radiation given by
Eq. (6) is

\begin{equation}
\rho_r \simeq \frac{m_{pl}^4 \lambda^4 \alpha}{2 \pi
(1+\alpha)^2}\,\left(\frac{\phi}{m_{pl}}\right)^2. 
\label{eq33}
\end{equation}

The end of inflation is achieved when $\epsilon_{wi} = 1$, or 

\begin{equation}
\left(\frac{\phi_{e}}{m_{pl}}\right)^2 \simeq \frac{1}{2 \pi
(1+\alpha)}\,\left(1+\frac{1}{\sqrt{1+\alpha}}\right). 
\label{eq34}
\end{equation}

\noindent This expression is valid either for  a
regime of very strong ($\alpha \gg 1$), or very weak dissipation ($\alpha \ll
1$). For the first, we obtain $\left(\frac{\phi_{e}}{m_{pl}}\right)^2 \simeq
\frac{1}{2 \pi \alpha}$, whereas for the second situation
$\left(\frac{\phi_{e}}{m_{pl}}\right)^2 \approx \frac{1}{\pi}$ as in the
super-cooled inflation. The beginning of warm inflation is indicated by $\phi =
\phi_*$ for which $T_r = H$. From Eqs. (\ref{eq4}) and (\ref{eq26}) it follows
that  

\begin{equation}
\left(\frac{\phi_{*}}{m_{pl}}\right)^2 \simeq \left(\frac{2.7 \alpha}{128 \pi^5
(1+\alpha)^2 \lambda^4}\right)^{1/3}.
\label{eq35}
\end{equation}

\noindent The next equation is 

\begin{equation}
N_{wi} = \frac{\pi (1 + \alpha)}{m_{pl}^2}\,(\phi^2_*-\phi_{e}^2) = 60,
\label{eq36}
\end{equation} 

\noindent obtained with Eqs. (\ref{eq4}) and (\ref{eq14}) and express the fact
that the total number of e-folds must be 
(at least) of the order of 60 to solve the usual problems mentioned in the 
introduction. These last two equations provide an important constraint 
between the free parameters $\alpha$ and $\lambda$. 

According to the discussion of Section III, we assume a small dissipation
($\alpha < 1$) such that we can neglect the contribution of the entropy
perturbations and the adiabatic perturbations will be relevant for structure
formation. However, we have no idea of how small actually $\alpha$ should be. A
possible way of making an estimative of this parameter is to use the
COBE-DMR\cite{cobe} observation of the cosmic microwave 
radiation anisotropy, which asserts for the temperature anisotropy
$\left(\frac{\Delta T}{T_0}\right)^2 \approx 3.43 \times 10^{-11}$, with $T_0 =
2.73 K$. Then, following the usual assumptions\cite{turner} concerning the
contribution of scalar perturbations to this measured quantity, we may have

\begin{equation}
\delta_{ad}^2 \simeq 3.43 \times 10^{-11},
\label{eq38}
\end{equation}

\noindent and this expression must be evaluated for a given scale crossing the
Hubble radius 50 e-folds before the end of inflation. We also remark 
after determining the value of $\alpha$ we can verify if at the end of
inflation there will be enough radiation necessary to the radiation era. 

As in super-cooled inflation (see appendix) a scale of astrophysical interest
crosses the Hubble radius at $N_{wi} = 50$. Thus, $\phi_{50}$ can be evaluated
from Eq. (\ref{eq36}), resulting 

\begin{equation}
\left(\frac{\phi_{50}}{m_{pl}}\right)^2 = \frac{50}{\pi (1 + \alpha)} + 
\left(\frac{\phi_{e}}{m_{pl}}\right)^2. 
\label{eq37}
\end{equation}

\noindent Now we are in conditions to determine the relevant parameters of the
model, namely, $\lambda$, $\alpha$, $\phi_*$ and $\phi_e$. For this, we
substitute Eqs. (\ref{eq31}) and (\ref{eq32}) into the expression for
$\delta_{ad}^2$ (Eq. (\ref{eq27})), and evaluate it at 50 e-folds before the
end of inflation through Eq. (\ref{eq37}). Finally, the last equation is
obtained when we take the observational data of COBE-DMR on the cosmic
microwave background anisotropy given by (\ref{eq38}).

Therefore, solving the system formed by Eqs. (\ref{eq34}), (\ref{eq35}),
(\ref{eq36}) and (\ref{eq38}), we find 

\begin{eqnarray}
& & \alpha \simeq 1.04 \times 10^{-9},\;\; \lambda^4 \simeq 0.97 \times 10^{-17}
\nonumber \\
& & \phi_* \simeq 4.40 m_{pl},\;\; \phi_{e} \simeq 0.56 m_{pl}
\label{eq39}
\end{eqnarray}

\noindent The temperature at the end of inflation is evaluated from the amount
of radiation at this epoch. Using the above parameters and Eq. (\ref{eq26}), it
is not difficult to arrive to the following temperature 

\begin{equation}
T_{end} \simeq 6.28 \times 10^{-8} m_{pl},
\label{eq40}
\end{equation}

\noindent which is in good agreement with the temperature generated by
reheating. These results  are quite interesting. In Refs. \cite{olivrud,leefang1}
the dissipation term is constrained in such a way to produce at the end of
inflation a hot universe, and consequently a smooth transition to the radiation
era. In these works previous estimations of the ratio between dissipation and the
Hubble parameter can be of order of $10^{-7}$ or less. Here, by considering
consistently the production of adiabatic perturbations and constraining the
amplitude of its spectrum with COBE data, we arrive to similar
results. Finally, the spectral index of scalar perturbations can be calculated
directly from Eq. (\ref{eq30}), yielding 

\begin{equation}
n_s \simeq 0.956
\label{eq41}
\end{equation} 

\noindent that is slightly close to one than the corresponding obtained for
super-cooled inflation that is found to be 0.941.

In this second and last  example we choose the quadratic potential
$V=\frac{1}{2} m^2 
\phi^2$ and a constant dissipation term, $\Gamma_0=constant$. During
inflation $\alpha$ is no longer constant given by

\begin{equation}
\alpha \simeq \frac{m_{pl} \Gamma_0}{\sqrt{12 \pi} m \phi}.
\label{eq42}
\end{equation}

\noindent The same steps adopted before will be followed straightforwardly
in order to estimate the values of $\Gamma_0$ and $m$ compatible with the
assumption of small dissipation and the observational data. The radiation is
assumed to be (see Eq. 4)

\begin{equation}
\rho_r \simeq \frac{\sqrt{3} \Gamma_0 m m_{pl} \phi}{96
\pi^{3/2}}\,\left(\frac{\phi}{m_{pl}} + \frac{\Gamma_0}{\sqrt{12
\pi} m}\right)^{-2}
\label{eq43} 
\end{equation}

\noindent from which the beginning of warm inflation can be calculated 

\begin{equation}
\frac{2.7 \Gamma_0}{16^2 \pi^5 \sqrt{12 \pi} m}\,\left(\frac{\phi_*}{m_{pl}}
+ \frac{\Gamma_0}{\sqrt{12 \pi} m} \right)^{-2} \simeq
\left(\frac{m}{m_{pl}}\right)^2\,\left(\frac{\phi_*}{m_{pl}}\right)^3.
\label{eq44}
\end{equation}

\noindent The end of inflation occurs when $\epsilon_{wi}=1$ or

\begin{equation}
\frac{1}{4 \pi}\,\left(\frac{m_{pl}}{\phi_e}\right)^2 + \frac{\Gamma_0
m_{pl}^3}{128 \sqrt{5}
\pi^{3/2} m \phi_e^3}\,\frac{\left(\frac{\phi_e}{m_{pl}} -
\frac{\Gamma_0}{\sqrt{12 \pi} m}\right)}{\left(\frac{\phi_e}{m_{pl}} +
\frac{\Gamma_0}{\sqrt{12 \pi} m}\right)^3} \simeq
\frac{m_{pl}}{\phi_e}\,\left(\frac{\phi_e}{m_{pl}} + 
\frac{\Gamma_0}{\sqrt{12 \pi} m}\right).
\label{eq441}
\end{equation}

\noindent The additional constraint comes from the imposition of $N_{wi} =
60$. then, Eq. (\ref{eq14}) for this case is

\begin{equation}
N_{wi} = 2 \pi\,\left[\left(\frac{\phi_*}{m_{pl}} +
\frac{\Gamma_0}{\sqrt{12 \pi}
m}\right)^2-\left(\frac{\phi_e}{m_{pl}} + 
\frac{\Gamma_0}{\sqrt{12 \pi} m}\right)^2\right] = 60
\label{eq45}
\end{equation}

\noindent The expression for $\phi_{50}$ is established form the
above by setting $N_{wi}=50$.

The last step is to express properly the spectrum of scalar perturbations for
this specific case in the same way as done before. Then, after a direct
calculation, taken into account the expression for $\phi_{50}$ and the
observational constraint (Eq. (\ref{eq38})), we obtained

\begin{eqnarray}
& & \Gamma_0 \simeq 3.09 \times 10^{-16}\,m_{pl},\;\; m \simeq 5.5 \times
10^{-8}\,m_{pl} \nonumber \\
& & \phi_* \simeq 3.09\,m_{pl},\;\; \phi_{e} \simeq 0.79\,m_{pl} 
\label{eq46}
\end{eqnarray}

\noindent Despite the very tiny value of $\Gamma_0$, it can be shown that, as
in the last example, $\alpha \propto 10^{-9}$ during warm inflation. Finally,
the temperature at the end of inflation is 

\begin{equation}
T_{end} \simeq 0.52 \times 10^{-7}\,m_{pl}
\label{eq47}
\end{equation}

\noindent which again is acceptable. The spectral index for adiabatic
perturbations is $n_s \simeq 0.972$, whereas the corresponding for super-cooled
inflation is $n_s \simeq 0.962$.

\section{Conclusions}

	In this paper we generalized the slow roll parameters applied in the 
usual super cooled inflation to the WI scenario. They are valid for any value of 
dissipation and can be used even if one wishes to allow time dependence of the 
energy density of radiation. Once we had well defined expansion parameters, we 
were able to study the evolution of the inflaton and radiation fields as well as 
of the entropy and adiabatic perturbations.
	
	We showed that the entropy perturbation $\delta S$ is proportional to 
the dissipation in the system. Since WI is feasible even for small values of 
$\Gamma$, we assumed that $\delta S$ is negligible, and focused on the 
subsequent evolution of the adiabatic perturbations. The restrictions imposed 
upon them by the observational data worked as a consistency check --- yielding 
small values for $\Gamma$ --- and provided the parameters (either $\lambda$ or 
$m$), final temperature and spectral index $n_s$ for the particular models 
chosen. Both presented a nearly scale-invariant spectrum, and temperatures 
comparable to the ones generated by the usual reheating mechanisms.
	
	As next steps, we will check if different potentials broadly adopted in 
WI generate self-consistent pictures as above. We also intend to study the 
behavior of tensor perturbations in the same scenario.

\section{Acknowledgments}

H. P. O. and S. E. J. acknowledge CNPq for financial support. We would like to
thank Dr. Kolb and Scott Dodelson for useful discussions. S. E. J. thanks
Robert Brandenberger for  his comments. S. E. J. is supported in Part by D.O.E. Grant 
(DE-FG02-91ER40688 - Task A). Finally, we have appreciated the comments of
A. Berera. 

\appendix

\section{Evolution equations for $\delta \phi$, $\delta \rho_r$ and $v$}

The total energy momentum
$T^{\mu}\,_{\nu} = T^{\mu}_{(r)\,\nu} + T^{\mu}_{(\phi)\,\nu}$, constituted by
radiation and scalar field, is conserved, or $T^{\mu\nu}\,_{;\nu}=0$. However,
radiation and scalar field are in interaction (cf. Eqs. (2) and (3)), whose
covariant description\cite{leefang2} is 

\begin{equation}
T^{\mu}_{(r)\,\nu}\,_{;\mu} = - T^{\mu}_{(\phi)\,\nu}\,_{;\mu} =
\Gamma\,(\phi_{,\alpha} U^{\alpha})\,\phi_{,\nu}
\end{equation} 

\noindent From the above equation we perturb the equation of energy balance of
radiation and scalar field to derive the evolution laws for
$\delta \rho_r$, $\delta \phi$, respectively, and also after perturbing the
equation of motion for the velocity field we obtain the evolution of $\delta
U_{\alpha}$ and consequently $v$. The calculations are tedious but direct and
result in 

\begin{eqnarray}
& &(\delta \rho_r \dot{)} + 4 H \delta \rho_r + \frac{4}{3} k a \rho_r v = 
4 \rho_r \dot{\Phi} + \dot{\phi}_0^2 \Gamma^{\prime}
\delta \phi + \Gamma \dot{\phi}_0\,(2 (\delta \phi\dot{)} - 3 \dot{\phi}_0
\Phi) \nonumber \\
& & (\delta \phi \ddot{)} + (3 H + \Gamma)\,(\delta \phi \dot{)} + \left(
\frac{k^2}{a^2} + V^{\prime\prime} + \dot{\phi}_0
\Gamma^{\prime}\right)\,\delta \phi = 4
\dot{\phi}_0 \dot{\Phi} + (\dot{\phi}_0 \Gamma - 2 V^{\prime})\, \Phi \\
& & \dot{v} + \frac{\Gamma \dot{\phi}_0^2}{\rho_r}\,v + \frac{k}{a}\,
\left(\Phi + \frac{\delta \rho_r}{4 \rho_r} + \frac{3 \Gamma \dot{\phi}_0}
{4 \rho_r}\,\delta
\phi \right) = 0 \nonumber
\end{eqnarray}

\noindent In order to obtain the leading term of the rhs of Eq. (\ref{eq22}), we
follow the approximation used in ref. \cite{plastaro}. Basically, it consists
in neglecting  $\dot{\Phi}$ and $(\delta \phi\ddot{)}$ during inflation, and in
addition we also neglect $\dot{v}$. Taking into account these approximations in
the Eq. (\ref{eq22}), we arrive to the following relation

\begin{equation}
\dot{\mathcal{R}} \simeq \frac{4 \Gamma \rho}{3 (\rho + p)}\,\left\{\Phi +
\frac{3 \sqrt{\pi \epsilon}}{2}\,\frac{\delta \phi}{m_{pl}} -
\frac{\epsilon}{12 (1+\alpha)}\,\left[3 \Phi + \left(\frac{3}{4} +
\frac{1}{1 + \alpha}\right)\,\frac{\delta \rho_r}{\rho_r}\right]\right\}.
\end{equation}

\noindent Since during inflation we can set $\epsilon \ll 1$, we may
approximate the above equation as given by Eq. (24). Nonetheless, even with a
more rigorous calculation it becomes clear that the entropy perturbations 
depends directly on the dissipation. 

\section{Matching equation in warm inflation}

The matching equation in warm inflation is easily obtained using the same
reasoning of standard super-cooled inflation. Then, a comoving scale
$k$ whose length $\lambda=\frac{1}{k}$ crosses the Hubble radius $H^{-1}$
during inflation when $a_k\,\lambda_k=H^{-1}$. The  matching equation relates
$k$ (or $\lambda$) with $N(k)$, the corresponding value of the number of 
e-folds before the end of inflation that this scale crosses the Hubble
radius. We can write

\begin{equation}
\frac{k}{k_{today}}=\frac{a_k H_k}{a_{today}
H_{today}}=\frac{a_k}{a_{end}}\,\frac{a_{end}}{a_{today}}\,
\frac{H_k}{H_{today}}
\end{equation}   

\noindent where the subscript `today' indicates the present values of the
quantity and $a_{end}$ is the scale factor at the end of inflation. Now,
$a_k=a_{end} e^{-N(k)}$. Note above that there is no reference to reheating. As
usual we assume adiabatic expansion since the end of inflation that implies
$\frac{a_{end}}{a_{today}} \simeq \frac{2.73 K}{T_{end}}$, where $T_{end}$ is
the temperature at the end of inflation, and its value is the same of the
temperature generated at the reheating. After some direct calculation we arrive
at

\begin{equation}
N(k) = 62 - ln\,\left(\frac{k}{a_{today} H_{today}}\right) - \frac{1}{4}
ln\,\left(\frac{\rho_{end}}{V_k}\right) -
ln\,\left(\frac{10^{16}\,GeV}{V_k^{1/4}}\right)
\end{equation}

\noindent where $T_{end}=\rho_{end}^{1/4}$. Now, the scales of astrophysical
interest today, say the size of a galaxy $\lambda \sim Mpc$ crosses, up to
logarithm corrections,  the outside the Hubble radius at $N_{gal} \sim 62
-4\,ln 10 \sim 50$, where $H^{-1}_{today} \sim 10^4\,Mpc$ and $a_{today} =
1$.

\end{document}